\documentclass[english,twocolumn]{emulateapj}
\usepackage{color}

\usepackage{epstopdf}
\usepackage{amsmath}
\makeatletter
\newcommand\lsim{\mathrel{\rlap{\lower4pt\hbox{\hskip1pt$\sim$}}
\raise1pt\hbox{$<$}}}
\newcommand\gsim{\mathrel{\rlap{\lower4pt\hbox{\hskip1pt$\sim$}}
\raise1pt\hbox{$>$}}} 
 
 \newcommand{\tot}{\mathrm{tot}}

\makeatother
\shorttitle{ Hot Jupiters in Binaries }
\shortauthors{Naoz et al.}
\makeatother
\begin{document}

\title{On the Formation of Hot Jupiters in Stellar Binaries}

\author{Smadar Naoz,\altaffilmark{1,2} Will M.\ Farr,\altaffilmark{2} and Frederic A.\ Rasio\altaffilmark{2,3}}

\altaffiltext{1}{Harvard Smithsonian Center for Astrophysics, Institute for
  Theory and Computation, 60 Garden St., Cambridge, MA 02138}
 
\altaffiltext{2}{Center for Interdisciplinary Exploration and Research in Astrophysics (CIERA), Northwestern University, Evanston, IL 60208, USA}

\altaffiltext{3}{Department of Physics and Astronomy, Northwestern University}


\begin{abstract}
  We study the production of Hot Jupiters (HJs) in stellar binaries.
  We show that the ``eccentric Kozai-Lidov'' (EKL) mechanism
  can play a key role in the dynamical evolution of a star--planet--star 
  triple system.  We run a large set of Monte Carlo
  simulations including the secular evolution of the orbits, general
  relativistic precession, and tides, and we determine the semi-major
  axis, eccentricity, inclination and spin--orbit angle distributions
  of the HJs that are produced. We explore the effect of different
  tidal friction parameters on the results.  We find that the
  efficiency of forming HJs when taking the EKL mechanism into account
  is higher then previously estimated.  Accounting for the frequency
  of stellar binaries, we find that this production mechanism can
  account for about $30\%$ of the observed HJ population.  Current
  observations of spin--orbit angles are consistent with this mechanism
  producing $\sim 30\%$ of all HJs, and up to $100\%$ of the
  misaligned systems.  Based on the properties of binaries without a
  HJ in our simulations, we predict the existence of many Jupiter-like
  planets with moderately eccentric and inclined orbits and
  semi-major axes of several AU.
\end{abstract}

\section{Introduction}\label{intro}

At least $\sim 20\%$ of exoplanets are associated with one or more
stellar companions
\citep{Raghavan+06,Desidera+07,Eggenberger+07,Mugrauer+09,Raghavan+10}.
Stellar companions may significantly alter the planetary orbits around
their partner on secular timescales.  
Close-in giant planets tend to be found preferentially in binary stellar systems
\citep{Zucker2002,Udry2007}. On the other hand recent studies suggest that the frequency of
giant planets in close binaries ($<100\,$AU) is significantly lower
then in the overall population \citep{Eggenberger+08,Eggenberger+11},
indicating that distant stellar perturbers may be important in the
production of HJs.

Recent measurements of the sky-projected angle between the orbits of
several HJs and the spins of their host stars have shown that
misalignment and even retrograde orbits are common
\citep[e.g.][]{GW07,Tri+10}.  If these planets migrated in from much
larger distances through their interaction with the protoplanetary
disk \citep{Lin+86,Mass+03}, their orbits should have low
eccentricities and inclinations \citep[but see][]{Lai+10,Thies+11}.
An alternative scenario that can account for the retrograde orbits involves
the secular interaction between a planet and a binary stellar
companion \citep{Wu+03,Dan,Wu+07,Takeda,Cor+11}.
 
Many theoretical studies have investigated the role of secular
perturbations in triple systems
\citep{Hol+97,Wu+03,Takeda05,Dan,Wu+07,Takeda,Cor+11,Naoz11,Naoz+11sec,Veras+12,KP12}. 
Long-term stability requires that the system be sufficiently hierarchical.  For
an inner binary (stellar mass $m_1$ and planet mass $m_2$) in a
nearly-Keplerian orbit with semi-major axis (SMA) $a_1$, the outer
orbit for mass $m_3$ around the center of mass of the inner binary
must have SMA $a_2 \gg a_1$. For stability the eccentricity of the
outer orbit, $e_2$, must also be small enough for $m_3$ to avoid close
approaches with the inner orbit.  In such systems a sufficiently
inclined $m_3$ can produce large-amplitude oscillations of the
eccentricities and inclinations of the $m_1$--$m_2$ system, while $a_1$
and $a_2$ remain nearly constant; this is the so-called 
Kozai-Lidov mechanism \citep{Kozai,Lidov}.

The standard treatment \citep{Kozai} assumes that $m_2 \ll m_1,\,m_3$, 
and $e_2 \simeq 0$.  Recently, \citet{Naoz11,Naoz+11sec} showed
that these approximations are not appropriate for many systems.  In
the presence of an eccentric outer orbit, or if the test particle
approximation for the inner planet is relaxed, the behavior of the system
can be qualitatively different.  The different behavior is associated
with breaking of the axial symmetry present in the standard Kozai
analysis; the associated lack of conservation of the axial component
of the inner orbit's angular momentum allows the orbit to reach
extremely high eccentricities and can even ``flip'' the orbit
from prograde to retrograde (with respect to the total angular
momentum).

Here we study the evolution of Jupiter-like planets in binaries
through Monte Carlo simulations.  We include an octupole-level
approximation to the perturbing potential and the tidal interactions
between the star and the planet \citep[following][]{Egg+01}, as well
as GR precession.

\begin{table*}[htb]
\begin{center}
  \begin{tabular}{|l c c c  || c c c|}
    \hline
    name          & $a_2$          & $e_1$     & $t_{V,2}$ & HJs  & $\psi>90^\circ$ & Total \\
                  & [AU]           & (initial) & [yr]      & $\%$ & $\%$            & Runs  \\
    \hline \hline
    SMA$1000$     & $100$          & $0.01$    & $1.5$     & $13$ & $50$            & 2478  \\
    SMA$100$      & $1000$         & $0.01$    & $1.5$     & $1$  & $60$            & 3165  \\
    SMA$500$      & $500$          & $0.01$    & $1.5$     & $19$ & $44$            & 2130  \\
    SMA$500$L     & $500$          & $0.01$    & $ 0.015$  & 18   & 46              & 2014  \\
    SMA$500$H     & $500$          & $0.01$    & $150$     & 0.5  & 25              & 2514  \\
    SMARan        & uniform in log & 0.01      & 1.5       & $16$ & $44$            & 4690  \\
    SMARane1R     & uniform in log & Rayleigh  & 1.5       & 10   & 39              & 1674  \\
    \hline
  \end{tabular}
\end{center}
\caption{\label{table_sim} Parameters of the Monte Carlo simulations.
  In all runs we start the Jupiter in a circular orbit at $5\,$AU. The 
  arguments of periastron, $g_1$ and $g_2$, were chosen from a uniform
  distribution, and the mutual inclination between the inner and outer
  orbits is drawn isotropically. On a 
  few days orbit, $t_{V}=1.5\,$yr is equivalent to $Q\simeq 2\times 10^5$
  while $t_{V}=0.015\,$yr and $t_{V}=150\,$yr (used in the SMA$500$L and
  SMA$500$H runs) are equivalent to $Q\simeq 2\times 10^3$ 
  and $2\times 10^7$, respectively.  For the star we choose
  $t_{V}=50\,$yr in all runs, which is equivalent to $Q \simeq 10^5$.}
\end{table*}

\section{Numerical Setup}\label{ICs}

We set $m_1=1\,M_\odot$, $m_2=1\,M_J$ and $m_3=1\,M_\odot$ for all our
calculations.  We denote the inclination angle of the inner (outer)
orbit with respect to the total angular momentum by $i_1$ ($i_2$), so
that the mutual inclination between the two orbits is
$i_\tot=i_1+i_2$.  We call the angle between the spin of the inner
star and the direction of the angular momentum of the inner orbit
$\psi$.  The projection of $\psi$ onto the plane of the sky,
$\lambda$, can be observed through the Rossiter--McLaughlin effect
\citep[e.g.,][]{GW07} and other methods
\citep[e.g.,][]{Nutzman+11,Sanchis+11,AviS+12}.

We solve the octupole-level secular equations numerically following
\citet{Naoz+11sec}. We are able to follow the spin vectors of both the
planet and the star.  The effects of tides and spins agree with those
of \citet{Dan} (private communication).  In many of our simulations,
the inner orbit reaches extremely high eccentricities. 
 During
excursions to high eccentricity there is a competition between the
increased efficiency of tides leading to the Kozai capture process
\citep{Naoz11} and the possibility of destroying the system by
crossing the Roche limit.

The upper limit for each system's integration time in all our
simulations was $9\,$Gyr.  However, if the planet becomes
tidally captured by the inner star, the integration becomes extremely
expensive.  We therefore stop the simulation, classifying the planet
as a HJ, whenever $a_1<0.03\,$AU and $e_1<0.01$, or $a_1<0.06\,$AU and
$e_1<0.4$. These conditions imply either a circular HJ, or an eccentric
one, respectively. If a planet crosses the Roche limit we also stop the 
run and we consider it lost (see below).
 
We draw outer periods from the log-normal distribution of
\citet{Duquennoy+91}, but require that the corresponding $a_2$ lies
between $51$ and $1500$~AU.  At these separations the distribution is
approximately constant in $\log (P_2)$, where $P_2$ is the outer orbit period.  The distribution of the outer
eccentricity is assumed thermal\footnote{Recently, \citet{Raghavan+10} showed that the eccentricity distribution for sun-like stars is closer to uniform. Therefore we have conducted an additional run similar to SMA$500$ but with the stellar-binary eccentricity drown from a uniform distribution. We have found that the results were essentially unaffected.}. 
We also performed runs for which we fix
$a_2=100$, $500$ or $1000$~AU.  These runs serve as a ``zoom-in'' since they give slices through the
larger SMA parameter space. Two additional runs have different
tidal friction parameters, varied over two orders of magnitude.  Motivated by
\citet{Moor+11}, we adopt a Rayleigh distribution for $e_1$ with a
mean eccentricity of $0.175$  for
one of the runs.  All runs are summarized in Table~\ref{table_sim}. 
Note that all our initial conditions are stable
according to the \citet{Mardling+01} criterion.

The differential equations that govern the inner binary's tidal
evolution were presented in \citet{Egg+01}. These equations take into
account stellar distortion due to tides and rotation, with tidal
dissipation based on the theory of \citet{1998EKH}.  The viscous time
scale, $t_{V}$, is related to the quality factor $Q$ \citep{GS66} by
\begin{equation}
Q=\frac{4}{3} \frac{k_{L}}{(1+2 k_{L})^2}\frac{G m}{R^3}\frac{t_{V}}{n} \ ,
\end{equation}
where $n=2 \pi/P$ is the mean motion of the orbit and $k_{L}$ is the
classical apsidal motion constant.  We use the typical value $k_{L} =
0.014$, valid for $n = 3$ polytropes, when representing stars and
$k_{L}=0.25$, valid for $n = 1$ polytropes, when representing gas
giant planets \citep{Egg+01,Dan}.  Table \ref{table_sim} specifies the
different viscous times we have used in our treatment for the
Jupiter-like planet.

The EKL mechanism can cause evolution of $e_1$ to extremely high
values, implying a high probability that a planet will cross the Roche
limit.  Following \citet{Soko} we define
\begin{equation}
R_L=\frac{R_2}{0.6} \left(\frac{m_2}{m_2+m_1} \right)^{-1/3} \ ,
\end{equation}
where $R_2=1\,R_J$. If we find that $a_1 (1-e_1) < R_L$ we stop the
run and assume that the planet is lost.

\begin{figure*}[htb]
\begin{center}
\plotone{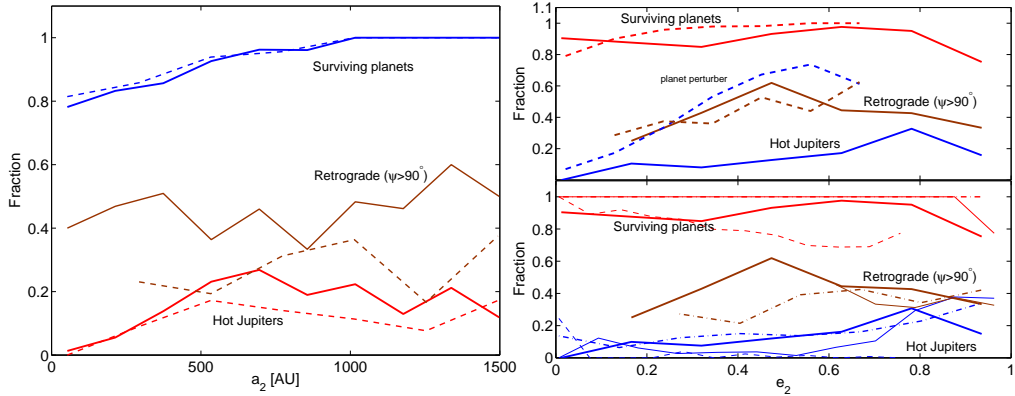}
\caption{Fraction of planets as a function of outer SMA (left) and the
  outer orbit eccentricity (right). In the left panel we consider the
  SMARan (solid lines) and SMARane1R (dashed lines) runs.  We show the
  fraction of systems that survive the EKL process (blue lines), the
  fraction of systems that formed HJs via Kozai capture (red lines)
  and the fraction of HJs that ended up in retrograde motion with respect
  to the stellar spin axis ($\psi>90^\circ$, brown lines).  In the
  bottom right panel we consider the SMA$500$ run (solid thick line),
  the SMA$1000$ run (solid thin line), the SMA$100$ run (thin dashed
  lines), and the run with extremely efficient tides, SMA$500$L
  (dot-dashed lines). The SMA$500$H run had extremely low efficiency
  of HJ formation and does not appear here. In the top right panel we
  compare the SMA$500$ run (solid thick line) to a system in
  which the perturber is another giant planet (dashed lines)
  \citep{Naoz11}.} \label{fig:sue}
\end{center}
\end{figure*}

\section{Results}\label{results}

\begin{figure}[htb]
\begin{center}
\plotone{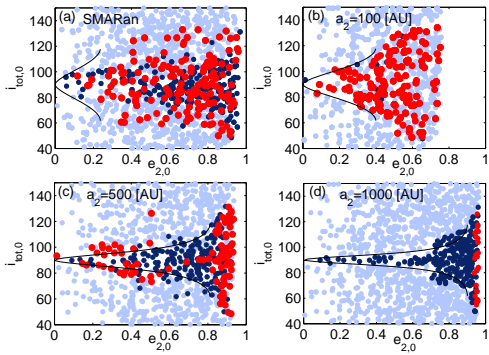}
\caption{Formation of HJs as a function of initial conditions. We
  consider the initial mutual inclination, $i_{\mathrm{tot},0}$, and
  the initial outer orbit eccentricity, $e_{2,0}$.  The complete suite
  of runs are marked by the light blue dots. In dark blue we mark the
  systems that form HJs, and in red we show the systems that did
  not survive the EKL process.  The solid line represents
  the analytical prediction for the first flip (see text). The
  different runs we considered are labeled in the
  figure. } \label{fig:sur}
\end{center}
\end{figure}

In Figure~\ref{fig:sue} we show the fraction of systems that have
survived the EKL mechanism as a function of SMA (for both the SMARan
and SMARane1R runs), and as a function of the outer orbit eccentricity
in the ``zoom-in'' runs.  In these figures we also show the fraction
of systems that formed HJs via the Kozai capture mechanism. For wide
separations ($a_2>100\,$AU) the EKL mechanism and Kozai capture are far
more efficient than suggested previously \citep{Wu+07,Dan}.  In Figure~\ref{fig:sue}, left panel, 
we also compare our results to those of \citet{Naoz11} for a
planetary perturber.  

In Figure~\ref{fig:sur} we show the initial conditions that are
associated with HJ formation.  Most of the systems that form HJs lie
within the bounds of an analytic criterion for determining the regions
of parameter space associated with flipping the inner orbit in the test particle approximation 
\citep{LN,Boaz2}.  Before the flip occurs, the inner orbit's
eccentricity becomes extremely high, often resulting in Kozai capture.
Overall, the EKL mechanism and Kozai capture produce HJs about $15\%$
of the time for the systems we considered.

\begin{figure*}
\begin{center}
\plotone{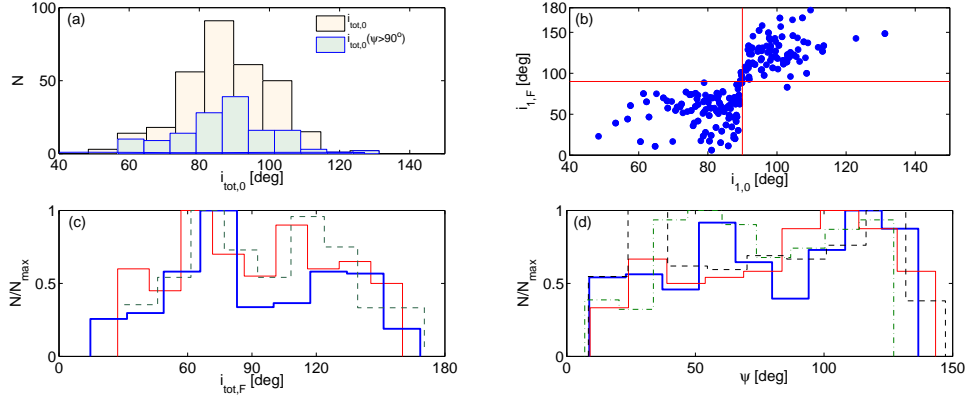}
\caption{Effect of initial conditions on the final state for HJs
  formed in the SMARan run.  In panel~(a) we show the distribution of
  the initial mutual inclination ($i_\tot$) for the systems that
  formed HJs. We also show in light colors the initial mutual
  inclination for those HJs that ended up in retrograde spin--orbit
  configuration (i.e., $\psi>90^\circ$). In panel~(b) we show a
  scatter-plot of the initial, $i_{1,0}$, and final,
  $i_{1,F}$, inner inclinations for the SMARan run. 
  Panel~(c) shows the distribution of the {\it final\/} mutual
  inclination for the SMARan run (thick solid blue line); we also show the
  distribution of the final mutual inclination for the SMA$500$ and
  SMA$1000$ runs (thin solid red and dashed black lines,
  respectively) as a function of the initial $i_{\tot,0}$.  Panel~(d) 
  shows the distribution of the
  final spin--orbit angle, $\psi$, for the SMARan run (thick solid blue
  line). We also show the distribution of the final $\psi$ for the
  SMA$500$, SMA$1000$ and SMA$500$L runs (thin solid red, dashed black
  and dot-dashed green lines, respectively). } \label{fig:incl}
\end{center}
\end{figure*}

In Figure~\ref{fig:incl} we show the distribution of mutual
inclination, $i_\mathrm{tot}$, and spin--orbit angle, $\psi$, resulting
from our simulations.  Consideration of the octupole-level secular
evolution when the inner body is a test particle \citep{Boaz2} shows
that the average part of the octupole Hamiltonian peaks for minimum
inner eccentricity of $\simeq 0.335$ and inclination $i_\tot\simeq 61.7^\circ$.  
This may be related to the peak observed in Figure~\ref{fig:incl} around $60^\circ$. 
An identical peak occurs for the
test particle octupolar Hamiltonian at the corresponding retrograde
inclination of $120^\circ$.  In this work we find that the retrograde
peak is wider, suggesting that retrograde orbits are more stable than
prograde orbits \citep{Innanen79,Innanen80,MG12}.


\section{Spin--Orbit Angles and Comparison with Observations}\label{retro}

Any proposed mechanism for producing misaligned HJs should be
consistent with both the observed \emph{shape} of the spin--orbit angle
distribution and the overall \emph{rate} of HJ formation.  Here we examine these questions
for HJs produced via the EKL mechanism and Kozai capture.

The efficiency of HJ formation from the EKL mechanism for close
($a_2\lsim 500\,$AU) binary systems is lower  for more distant
perturbers, in accordance with observations
\citep[e.g.,][]{Eggenberger+08,Eggenberger+11}.  The formation
efficiency for wide binaries is almost independent of both SMA
and eccentricity for the outer orbit (see Figure~\ref{fig:sue}). 
Following \citet{Wu+07} we estimate the fraction of
stars with HJs as
\begin{equation}
  f = f_b f_p f_\mathrm{EKL},
\end{equation}
where $f_b$ is the fraction of stars in binary systems, $f_p$ is the
fraction of stars with Jupiter-mass planets formed at a few AU, and
$f_{\rm EKL}\simeq 0.15$ is the efficiency of our mechanism from simulations.
Taking $f_b\simeq 0.3$, estimated from the tail of the distribution
in \citet{Duquennoy+91}, and $f_p\simeq 0.07$, consistent with the
studies from \citet{Wright+12} and \citet{Marcy+05}, we find $f\simeq 0.0032$.  
The fraction of stars hosting HJs is about $1\%$, so
we estimate that our mechanism can provide $\sim 30\%$ of the total HJ formation.

\begin{figure*}[htb]
  \plotone{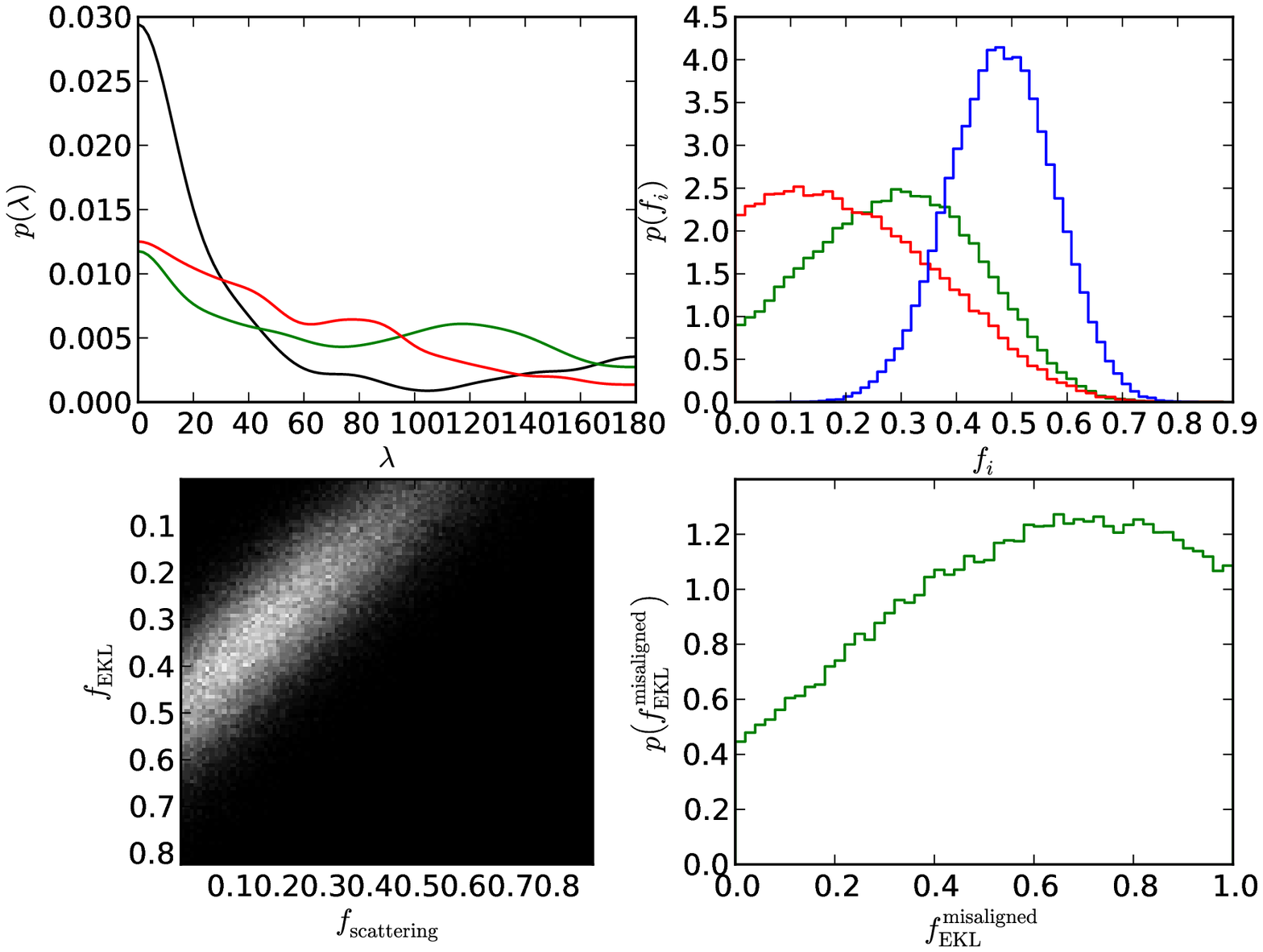}
  \caption{\label{fig:admixture} (Upper Left) Distributions of
    projected (i.e., observed) spin--orbit angle, $\lambda$: in black,
    the observed distribution of spin--orbit angles in the 45 systems
    of \citet{Exoplanetsorg} (the distribution has been smoothed using
    a Gaussian kernel that appropriately takes into account the
    observational errors in each system); in green, the distribution
    of spin--orbit angles produced by the EKL mechanism discussed in
    this work; and in red, the distribution of spin--orbit angles
    produced by the planet--planet scattering simulations in
    \citet{Nag+11}.  (Upper Right) Posterior distributions of the
    fractional contribution of the various components to the overall
    spin--orbit angle distribution implied by the data (see 
    eq.~\eqref{eq:spin-orbit-components} and~\eqref{eq:Bayes}): in green,
    distribution of the fraction of the spin--orbit distribution due to
    the EKL mechanism ($f_\mathrm{EKL}$) implied by the data; and in
    red the distribution of the fraction due to planet--planet
    scattering \citep[$f_\mathrm{scattering}$;][]{Nag+11}; and in blue
    the fraction attributed to disk migration ($1- f_\mathrm{EKL} -
    f_\mathrm{scattering}$).  (Lower Left) Posterior probability
    distribution in the $f_\mathrm{scattering}$-$f_\mathrm{EKL}$
    plane.  (Lower Right) Probability distribution for the fraction of
    misaligned systems produced by EKL (see eq.~\eqref{eq:fmisaligned}). \vskip 1cm }
\end{figure*}

Figure~\ref{fig:admixture} shows the observed \emph{projected}
spin--orbit angle distribution, the distribution produced by our
mechanism, and the distribution produced by the dynamical
planet--planet scattering considered in \citet{Nag+11}\footnote{See also the recent calculations by \citet{Boley+12} and \citet{BN}.}.
It is clear
that no one mechanism is a good fit to the observed distribution: more
systems are observed to be aligned than are produced by any of the proposed
formation mechanisms.  Aligned systems are a more natural consequence
of disk migration \citep{Lin+86,Mass+03}.  Following \citet{MJ}, we express
the complete spin--orbit angle distribution as a sum of contributions
from an aligned component, an EKL component and a dynamical
planet--planet scattering component \citep{Nag+11}
\begin{eqnarray}
  \label{eq:spin-orbit-components}
  p(\lambda | \left\{ f_i \right\}) &=& \left( 1 - f_\mathrm{EKL} - f_\mathrm{scattering}\right) \delta(\lambda)  \\
  &+&  f_\mathrm{EKL}\,
  p_\mathrm{EKL}(\lambda) + f_\mathrm{scattering}\,
  p_\mathrm{scattering} (\lambda) \ , \nonumber
\end{eqnarray}
with $\delta$ the Dirac delta function and $p_\mathrm{EKL}(\lambda)$
and $p_\mathrm{scattering} (\lambda)$ the distributions shown in
Figure~\ref{fig:admixture}.  The $f_i$ represent the relative
contribution of each component to the total distribution.  We impose
uniform priors on the $f_i$ with $f_i \geq 0$ and $\sum_i f_i \leq 1$,
and compute the posterior probability of the set of $f_i$, $\left\{
  f_i \right\}$, given the observed spin--orbit alignments (and
associated errors) of 45 systems%
\footnote{The spin--orbit observations reported at
  \url{http://exoplanets.org} \citep{Exoplanetsorg}, 
   include data from
  \citet{Simpson2011,Narita2010,Tri+10,Queloz2010,Guenther2012,Triaud2009,Winn2009,Johnson2009,Jenkins2010,Moutou2011,Wolf2007,Gillon2009,Winn2011,Winn2005,Brown2012,Winn2007,Winn2010,Johnson2008,Hellier2011,Bouchy2008,Narita2009,Hebrard2011,Simpson2010,Anderson2011,Narita2007,Hebrard2010,Pont2010}.} %
from \citet{Exoplanetsorg}, $d$, using Bayes rule:
\begin{equation}
  \label{eq:Bayes}
  p\left( \left\{ f_i \right\} | d \right) \propto p\left( d | \left\{
      f_i \right\} \right) p\left(\left\{ f_i \right\} \right).
\end{equation}
Treating each spin--orbit angle measurement as a Gaussian with errors given
by \citet{Exoplanetsorg}, the likelihood for the data set, $d$, is
given by
\begin{equation}
  \label{eq:likelihood}
  p\left( d | \left\{f_i \right\} \right) = \prod_{i=1}^{45} \int
  d\lambda \, N\left(\lambda; \lambda_i, \sigma^\lambda_i\right)
  p\left(\lambda | \left\{f_i\right\}\right), 
\end{equation}
where the $\lambda_i$ are the observed angles,
$\sigma_i^\lambda$ the observational errors, and $N(x; \mu, \sigma)$
is the Gaussian PDF with mean $\mu$ and standard deviation $\sigma$
evaluated at $x$.  The integral is evaluated over the range $\lambda
\in [0, \pi]$, properly accounting for wrapping at the endpoints.  The
resulting probability distributions%
\footnote{This analysis was performed using customized code
  implementing the algorithm described in \citet{Foreman-Mackey2012}
  and references therein.} %
for the fractional contribution of each component (aligned, EKL, and
planet--planet scattering) appear in Figure~\ref{fig:admixture}.  We
also show in Figure~\ref{fig:admixture} the probability distribution
for the EKL contribution to the misaligned systems,
\begin{equation}
  \label{eq:fmisaligned}
  f^\mathrm{misaligned}_\mathrm{EKL} \equiv
  \frac{f_\mathrm{EKL}}{f_\mathrm{EKL} + f_\mathrm{scattering}}
\end{equation}

As expected from the distributions in Figure \ref{fig:admixture}, the
data support a significant aligned component, accounting for about
50\% of the observed systems.  The posterior is nearly degenerate
along the line $f_\mathrm{scattering} + f_\mathrm{EKL} \simeq 0.5$
corresponding to a total contribution from scattering and EKL of 50\%.
However, the data do prefer a larger contribution from EKL than
scattering, with EKL accounting for most likely $\simeq 30\%$ of the
observed systems and planet--planet scattering $\simeq 10\%$ to
20\% of the systems.  The data prefer that EKL produces between
$60\%$ and $80\%$ of the misaligned systems, but fractions as low as
$0\%$ and as high as 100\% cannot be ruled out. Note that the allowed 
contribution from EKL
to the shape of the spin--orbit angle distribution is consistent with
the estimate from the rate of HJ production that EKL can account for
about $30\%$ of all HJ systems.

Previous studies \citep{Dan} of the secular effects of stellar
perturbers have considered only the quadrupole terms in the secular
potential.  The expansion up to octupole order used in our work
produces qualitatively different behavior.  The additional terms in
the potential can drive the inner orbit to much more extreme
eccentricities and inclinations (including retrograde inclinations,
which are impossible in the quadrupole limit), and, based on Figure~\ref{fig:sue}, 
leads to much more efficient HJ formation than found in
\citet{Dan}.

\section{Summary and Discussion}\label{dis}

We studied the formation and evolution of HJs in wide stellar
binaries using an approximation accurate to octupole order in the SMA
ratio.
Recent studies have shown that octuple-level perturbations can play a
very important role in the dynamics of three-body systems
\citep{Naoz+11sec}. \citet{Naoz11} showed that in the presence of a
second planetary-mass perturber, secular perturbations can easily produce
retrograde orbits. 

From an observational point of view, statistical analyses suggest that
at least $\sim 20\%$ of the known extrasolar planetary systems are
associated with one or more stellar companions
\citep{Raghavan+06,Desidera+07,Eggenberger+07}, and thus stellar
perturbers are likely to be at least as important as planetary
perturbers.

Our results differ from those of previous studies \citep{Dan,Wu+07} conducted at
quadrupole order.  We have found that the EKL mechanism produces HJs
through Kozai capture about $15\%$ of the time.  Given realistic assumptions 
about the rate
of binary stellar systems and the fraction of systems that initially
host planets, we find that our mechanism can produce HJs in about $0.3\%$ of
stars.  Since about $1\%$ of stars host a HJ, EKL may account for
about $30\%$ of all HJs.

By comparing the shape of the sky-projected spin--orbit angle
distribution produced by EKL, dynamical planet--planet scattering
\citep{Nag+11}, and disk migration, to the observed spin--orbit angle
distribution, we find that EKL likely contributes to about 30\%
of the observed distribution, consistent with the overall rate of HJ
production.

\section*{Acknowledgments}
This research was supported in part by NASA Grant NNX12AI86G, and through
the computational resources and staff contributions provided by
Information Technology at Northwestern University as part of its
shared cluster program, Quest. We thank Yoram Lithwick for the use 
of his time allocation on Quest. This research made use of the
Exoplanet Orbit Database and the Exoplanet Data Explorer at
exoplanets.org.

\bibliographystyle{hapj}

\bibliography{Kozai}
\end{document}